\documentclass[a4paper]{aa}
\usepackage{txfonts,graphicx,natbib}
\bibpunct{(}{)}{;}{a}{}{,}
\newcommand{\pv}{\ensuremath{P_V}}
\newcommand{\nv}{\ensuremath{N_V}}
\newcommand{\Bd}{\ensuremath{B_\mathrm{d}}}
\newcommand{\Bdo}{\ensuremath{B_\mathrm{d}^{(0)}}}
\newcommand{\Bq}{\ensuremath{B_\mathrm{q}}}

\newcommand{\Bz}{\ensuremath{\langle B_z\rangle}}

\newcommand{\cz}{\ensuremath{C_z}}
\begin{document}
\title{Observational constraints on the magnetic field of RR Lyrae stars}
       \author{
        Katrien Kolenberg \inst{1}
       \and
        Stefano Bagnulo   \inst{2}
        }
\institute{Institut fuer Astronomie, 
           Universitaet Wien,
           Tuerkenschanzstrasse 17,
           A-1018 Wien, Austria.\\
           \email{kolenberg@astro.univie.ac.at}
           \and
           Armagh Observatory,
           College Hill,
           Armagh BT61 9DG,
           U.K.
           \email{sba@arm.ac.uk}
}
\authorrunning{K. Kolenberg \& S. Bagnulo}
\titlerunning{Constraints on the magnetic field of RR Lyrae stars}

\date{Received: 2008-12-24 / Accepted: 2009-01-30}
\abstract
{
A high percentage of the astrophysically important RR Lyrae stars
show a periodic amplitude and/or phase modulation of their pulsation
cycles. More than a century after its discovery, this ``Blazhko effect''
still lacks acceptable theoretical understanding. In one of the plausible
models for explaining the phenomenon, the modulation is caused by the effects of a magnetic field. 
So far, the available observational data
have not allowed us to either support nor rule out the presence of a
magnetic field in RR Lyrae stars.
}
{
We intend to determine whether RR~Lyrae stars are generally characterized
by the presence of a magnetic field organized on a large scale.
}
{ 
With the help of the FORS1 instrument at the ESO VLT we performed
a spectropolarimetric survey of 17 relatively bright southern RR
Lyrae stars, both Blazhko stars and non-modulated stars, and determined
their mean longitudinal magnetic field with a typical error bar $<30$\,G.
}
{
All our measurements of the mean longitudinal magnetic field resulted in
null detections within 3\,$\sigma$.  From our data we can set an upper limit for the
strength of the dipole component of the magnetic fields of RR Lyrae
stars to $\sim 130$\,G. Because of the limitations intrinsic to the
diagnostic technique, we cannot exclude the presence of higher order
multipolar components.
}
{
The outcome of this survey clarifies that the Blazhko modulation in
the pulsation of RR Lyrae stars is not correlated with the presence of
a strong, quasi-dipolar magnetic field.
}


\maketitle
\section{Introduction}
Investigations of the pulsating RR Lyrae stars contribute to almost
every branch of modern astronomy. These stars serve as standard
candles to fix the cosmological distance scale and are often
considered to be prototypes of purely radial pulsation.  With mean
periods of about half a day and brightness variations of about one
magnitude, RR~Lyrae stars pulsate in the radial fundamental mode (type
RRab), the radial first overtone (type RRc), and both of these radial
modes simultaneously (type RRd). However, additional cycles occur in
many RR Lyrae stars. A considerable fraction of RR Lyrae stars
\citep[20-30\,\% of Galactic RRab and 5\,\% of RRc
stars;][]{MosPor03}, or even higher numbers
(about 50\% according to the outcome of the Konkoly Blazhko survey - Jurcsik, private communication),
shows a periodic amplitude/phase modulation on
timescales of typically tens to hundreds of times the pulsation
period. This phenomenon is denoted \textit{Blazhko effect}
\citep{Blazhko07}.  More than a century after its discovery, the Blazhko
effect remains a mystery.  The most commonly stated hypotheses to
explain the phenomenon invoke intrinsic effects, such as resonance,
a magnetic field,
or variable turbulent 
convection.

In the resonance models the modulation is caused by a (nonlinear) resonance between the
radial fundamental mode and a nonradial mode, most likely a dipole $(\ell = 1)$
mode \citep{Cox93,vHoletal98,DziMiz04}.   Like the
oblique pulsator model for roAp stars \citep{Kurtz82}, the magnetic models suppose
that Blazhko stars have a magnetic field inclined to the stellar
rotation axis \citep{Cousens83,ShiTak95}. The magnetic field, assumed to be a dipole field, deforms the radial mode leading to an additional quadrupole
component $(\ell = 2)$, for which the symmetry axis coincides with the
magnetic axis. With the star's rotation, our view of the pulsation
components changes, causing the observed amplitude modulation. According to this model, 
a magnetic dipole field with a strength of about 1\,kG is needed for the
amplitude modulation to be observable \citep{ShiTak95}.
Both the resonance and the magnetic models involve nonradial pulsation components, 
of which the presence in RR Lyrae stars still remains to be proven \citep{Chaetal99,Kolenberg02}.  
An alternative scenario has recently been proposed by \citet{Stothers06}.
 This model is based on purely radial pulsation and involves 
 a cyclic variation of turbulent convection 
 in the hydrogen and helium ionization zones of Blazhko stars.  According to \citet{Stothers06}, 
 the variation of the turbulent convection could be caused by a transient magnetic field in the star (turbulent dynamo mechanism). 
The expected strength of such a field is not mentioned, and it would be harder to detect than a large-scale dipole field.

Photometric observations often hint at a much more complex scenario 
for Blazhko stars than a modulation with one single period.  The
prototype RR~Lyr shows a longer cycle of about 4 years, beside
its primary 0.567\,d pulsation period, and its 40\,d Blazhko period
\citep[which has decreased over the past decades --
see][]{Koletal06}. At the end of this cycle, the Blazhko effect
suddenly weakens in strength and reappears with a change in the
Blazhko phase \citep{DetSze73}. \citet{Szeidl76} also reports
stronger and weaker 4-year cycles in the Blazhko effect of RR
Lyr. Other well-studied stars also reveal such longer cycles: about
3540\,d in XZ Cyg \citep{LaCluyze04}, about 7200\,d in XZ Dra
\citep{Juretal02}, and between 2000 and 3000\,d in RV UMa
\citep{Kovacs95}.  In addition, extensive analyses of old and new photometry (same
references as above) show that these well-studied stars exhibit
coincident changes of both the primary pulsation period and the
Blazhko period.  Some Blazhko stars are also known to have two 
modulation periods, such as XZ Cyg \citep{LaCluyze04}
and UZ UMa \citep{Sodetal06}.
These photometric observations clash with the models
requiring the Blazhko period to be exactly equal or directly
proportional to the rotation period of the star, such as the latest
version of the magnetic model \citep{Shi2000}.  However, long-term
cyclic changes could still be interpreted in terms of the magnetic
rotator-pulsator model, by explaining the observed phenomena with
changes of the global magnetic field structure and/or strength.

From the observational point of view, the question whether RR~Lyrae
stars are magnetic is still a matter of debate. Until recently, high-precision 
spectropolarimetric measurements of RR Lyrae stars were hampered by 
the brightness requirements of the technique.
RR Lyr, the prototype of the class, is by far the brightest
with $V=7.2-8.2$, and it is the only object of its
kind that has been the target of spectropolarimetric observations, the
outcome of which is contradictory. \citet{Babcock58} and
\citet{Rometal87} reported a
variable magnetic field in RR Lyr with a strength up to 1.5\,kG.
\citet{Preston67} and \citet{Chaetal04} reported null detections. The
last group of authors obtained magnetic field measurements of RR~Lyr with the
MuSiCoS spectropolarimeter attached to the 2\,m telescope at Pic du
Midi (France) from 1999-2002.  Having covered various phases in the
pulsation and the Blazhko cycle, they concluded from their data
($\sigma_B \approx 80$\,G) that RR Lyr is a bona fide non-magnetic
star. 

Though no significant dipole field was detected in RR Lyr by
\citet{Chaetal04}, there is the need for a more exhaustive test of the magnetic field hypothesis to explain the
Blazhko phenomenon, which can be performed through a survey of a
sample of RR~Lyrae stars.  To fully strengthen or disprove the
hypothesis of a magnetic field being the driving force behind the
Blazhko effect, we have a strong case by also checking non-modulated
stars.  A crucial test of the magnetic models for the Blazhko effect
would be to investigate the presence of magnetic fields in a sample of
RR Lyrae stars with different pulsational properties.  Thanks to the
highly efficient instrument FORS1 attached to VLT unit Kueyen,
numerous RR~Lyrae stars have now come within reach for
spectropolarimetry. This paper presents the first survey of magnetic
fields in a sample of RR~Lyrae stars.

In Section\,2 we discuss the targets, the observations and their reduction.
The determination of the mean longitudinal magnetic field is described in 
Section\,3. The outcome of our survey, in terms of the constraints on the 
magnetic field and the implications for the models for the Blazhko effect, 
is discussed in Section\,4, and concluding remarks are given in Section\,5.

\section{Observations and data reduction}

\subsection{Target selection}
We performed a high precision ($\sigma_B \la 30$\,G) survey for
magnetic fields in a sample consisting of 17 RR Lyrae stars in the
southern hemisphere.  The sample includes  4 non-modulated RRab (fundamental mode pulsators),
2 non-modulated RRc stars (first overtone pulsators), 10 Blazhko RRab stars, and 1
Blazhko RRc star. All of the
targets are case studies in the understanding of RR Lyrae pulsation.
The question we wanted to address is which group(s) (if any) of these
stars show(s) a detectable magnetic field, and in particular if a
magnetic field is detected in the Blazhko stars \citep[see
also][]{Koletal03}. For comparison, we obtained one spectrum of a magnetic A star (HD~94660).
Our sample of stars is given in Table~\ref{Tab_Sample}.
Magnitude ranges are taken from the GCVS \citep[General Catalogue of Variable Stars][]{Kholopov99}, and
periods from the GCVS, the GEOS RR Lyrae
database, \citet{LeBoretal08} or Kolenberg et al. (2007, 2008).

The magnetic field strength may vary over the pulsation cycle and the
Blazhko cycle, as was reported by \citet{Rometal87}.  Based on the
times of maxima given in the GEOS RR Lyrae database ({\tt
http://rr-lyr.ast.obs-mip.fr/dbrr/}), which contains also recent TAROT
observations \citep[][and references therein]{LeBoretal08}, and from
our own data \citep{Koletal07,Koletal08}, we were able to determine
the pulsation phases of our observed stars with a precision of 0.02.
For the determination of the Blazhko phases, in contrast, we generally
(except for SS For) could not rely on published accurate Blazhko
ephemerides.  In order to determine the Blazhko ephemerides for the
observed stars, we used the parameters published by
\citet{WilSod05} to fit the available data of the stars, ASAS data as
well as our own measurements (if available).  Monte Carlo simulations
were used to determine the error on the Blazhko period.  From the
available photometric data we then extrapolated the Blazhko phase at the
time of our FORS1 measurements. The error on the phase determination is
dominated by the contribution of the error on the Blazhko period
multiplied by the number of Blazhko cycles elapsed from the time of
photometric measurements and of magnetic measurements. For all stars
of our sample, the error in the determination of the Blazhko phase
turned to be substantially higher than for pulsation phase. For some
stars, those for which the photometric data were taken a few years
prior to the FORS1 measurements, the uncertainty on the Blazhko phase
turned out to be higher than 0.5, in which case our phase estimate is
nearly meaningless. 

Having 2.5 nights of observing time at our disposal in visitor mode, we optimized the 
observing strategy (number of different targets, slit width and exposure time) to 
get the most stringent constraints on the magnetic field (see Section\,4.2).

\begin{table*}
\caption{\label{Tab_Sample}
Our sample of RR Lyrae stars for the spectropolarimetry and the
magnetic A star for comparison.
}
\begin{center}
\begin{tabular}{lllcl}
\hline \hline
\multicolumn{2}{c}{Object}     &
\multicolumn{1}{l}{Type}       &
\multicolumn{1}{c}{$V$ (mag)}  &
\multicolumn{1}{c}{$P$ (d)}   \\
\hline
        &              &        &                   &          \\
WZ Hya  & HIC 50073   & RRab    & 10.27-11.28 &  0.53772  \\ 
V Ind   & HIC 104613  & RRab    &  9.12-10.48 & 0.47959  \\ 
U Lep   & HIC 22952   & RRab    &  9.84-11.11 &  0.58148  \\ 
\vspace{2mm}
RU Scl  & HIC 226     & RRab    &  9.35-10.75 &  0.49332  \\
CS Eri  & HIC 12199   & RRc     &  8.75- 9.31 &  0.31133  \\ 
\vspace{2mm}
MT Tel  &  HIC 93476  & RRc     &  8.68- 9.28 &  0.31690  \\ 
RV Cap  & HIC 103755  & RRab-BL & 10.22-11.57 &  0.44774  \\ 
RX Cet  & HIC 2655    & RRab-BL & 11.01-11.75 &  0.57369  \\ 
RV Cet  & HIC 10491   & RRab-BL & 10.35-11.22 &  0.62340  \\ 
RX Col  & HIC 29528   & RRab-BL & 11.4 -12.6  &  0.59404  \\ 
RY Col  & HIC 24471   & RRab-BL & 10.44-11.24 &  0.47884  \\ 
VW Dor  & HIC 29055   & RRab-BL & 11.22-12.11 & 0.57061  \\ 
SS For  & HIC 9932    & RRab-BL &  9.45-10.6  &  0.49543  \\ 
RX For  & GSC6442.00690& RRab-BL& 11.12-12.46 &  0.59731  \\
SZ Hya  & HIC 45292   & RRab-BL & 10.44-11.84 &  0.53724  \\
\vspace{2mm}
X Ret   & LB 3316     & RRab-BL & 11.16-12.14 &  0.49199  \\ 
\vspace{2mm}
BV Aqr  & HIC 108839  & RRc-BL  & 10.8 -11.2  &  0.36405  \\ 
\vspace{1mm}
HD~94660 & HIC 53379   & Magnetic Ap & 6.09    &  -        \\ 
\hline
\end{tabular}
\end{center}
\end{table*}

\subsection{Spectropolarimetry with FORS1}
Observations were obtained in visitor mode during nights 2008-11-10 to
2008-11-13 with the Focal Reducer/Low Dispersion Spectrograph FORS1
\citep{Appetal98} of the ESO Very Large Telescope (VLT). We used
grism 1200\,B, which covers the spectral
range of $\sim 3700-5100$\,\AA\ with a dispersion of 0.36\,\AA\ per
pixel.  Slit width was set either to 0.5\arcsec\ or 1.0\arcsec,
according to seeing conditions, for a spectral resolution of about
2800 and 1400, respectively. CCD readout mode was set to 200\,kHz,
with a $2 \times 2$ binning (for a pixel spatial scale of 0.25\arcsec) and
with a gain corresponding to $\sim 2.2\,$e$^-$/ADU.

The instrument was used in spectropolarimetric mode. Circular
polarization measurements were performed after insterting a
quarter-wave plate (at set position angles) and a Wollaston prism into
the parallel section of the instrument's optical path. For each star, we
obtained eight exposures with the position angle of the quarter-wave 
plate set at $-45\degr$, $+45\degr$, $+45\degr$, $-45\degr$,
$-45\degr$, $+45\degr$, $+45\degr$, $-45\degr$. Such a high number of
exposures was necessary to obtain a very high signal-to-noise ratio
(SNR) without saturating the CCD. Setting the retarder waveplate at
two different position angles allowed us to minimize the contributions
of spurious (instrumental) polarization \citep[e.g.][]{Semetal93}.
The signal-to-noise ratio cumulated over both beams and all retarder
waveplate positions, estimated in the wavelength range 4475-4525\,\AA,
was between 1800 and 3000 per \AA, allowing us to reach a typical
error bar of 0.05\,\% in the \pv\ (the ratio
between Stokes $V$ and Stokes $I$) profile in a 1\,\AA\ spectral bin.

Data were reduced using standard IRAF routines and a dedicated FORTRAN
code. All the science frames were bias subtracted using a master bias
obtained from a series of five frames taken the morning after the
observations.  No flat fielding procedure was applied to the data.
Spectrum extraction was performed by collapsing a 60\,pixel
($=15\arcsec$) wide strip centred about the traced central peak. The
extraction parameters were obtained from the first exposure of each
series, and then adopted for all the remaining exposures, so as to
minimize the effects of a different sensitivity of the CCD between the
pixel regions illuminated by the two beams split by the Wollaston. Sky
subtraction was performed selecting symmetric regions on the left and
right side of each spectrum (typically between pixel 30 and 35 from
the central peak), and fitting those with a Chebyshev polynomial. 
Owing to the limited size of the CCD region illuminated by each
beam split by the Wollaston prism, it was not possible to calculate
the sky contribution using a wider region.  In fact,
since our targets are relatively bright, sky subtraction is not
a critical step, and could even have been skipped without significantly
affecting the final results.

Wavelength calibration was based on one arc frame obtained the morning
after the observations, and typically led to a RMS scatter of $\la
0.02$\,pixels. Fine-tuning of wavelength calibration based on
night sky lines could not be performed, therefore the accuracy of the
absolute wavelength calibration is restricted by instrument flexures,
which are expected to be less than 1 pixel up to a zenith distance of
60\degr\ (see FORS1/2 User Manual).

The final products of data reduction are the \pv\ profiles (the ratio
between Stokes $V$ and Stokes $I$), together with their error bars,
calculated as explained in \citet{Bagetal06}, i.e., combining all
eight exposures filtered using a $\sigma$-clipping algorithm. In
addition, we obtained the so-called ``null profiles'', \nv\
profiles \citep{Donetal97}, which are expected to be distributed around
0 with the same FWHM that characterizes the \pv\ profiles. A deviation
from 0 of the \nv\ profile exceeding $3\,\sigma$ would prompt for a
re-inspection of the data and data reduction quality, and for a search
for potential causes of spurious polarization, such as, e.g., those
due to spectral variability during the exposures.

It is well known that in RR Lyrae envelopes, especially those of
RRab type stars, strong acoustic waves occur, and shock waves are produced at
certain pulsational phases \citep[][and references
therein for recent observations]{Presetal65,Chaetal08}. The gas dynamics associated with these waves in
the line forming region, as well as the existence of the preheating
zones ahead of the shock fronts and the cooling zones behind them,
can strongly affect the shape of the spectral lines in RR Lyrae
stars, resulting in line asymmetry, additional broadening, line
profile doubling, and emission components. These distortion effects
are particularly present during the rise to maximum light, around
the phase of minimum radius.

As pointed out by \citet{Wadetal02} and \citet{Chaetal04}, the rapid
changes in the line profiles of both the Balmer and the metallic lines
may lead to spurious polarization signals in high-resolution
spectropolarimetric data.  However, it is reasonable to assume that
this problem is less critical with the technique based on
low-resolution spectra adopted in this work.  Indeed, at a spectral
resolution of 1400 or 2800, as was obtained with FORS1 for this study,
most metal lines are not resolved, let alone their variations.  Most of
our observations were obtained outside of the phase range in which the
stars are on the steep rising branch of their light curve (typically
the pulsation phase interval $\phi=0.9-1.0$).  For the few stars in
our sample of which we obtained spectra close to the shock phase, we
could not see any effects on the spectra, in particular no spikes show
up in the null profiles, as we would expect if the resolved line profiles were
significantly changing during the observing time.

\begin{figure*}[ht]
\begin{center}
\includegraphics[width=130mm,angle=270]{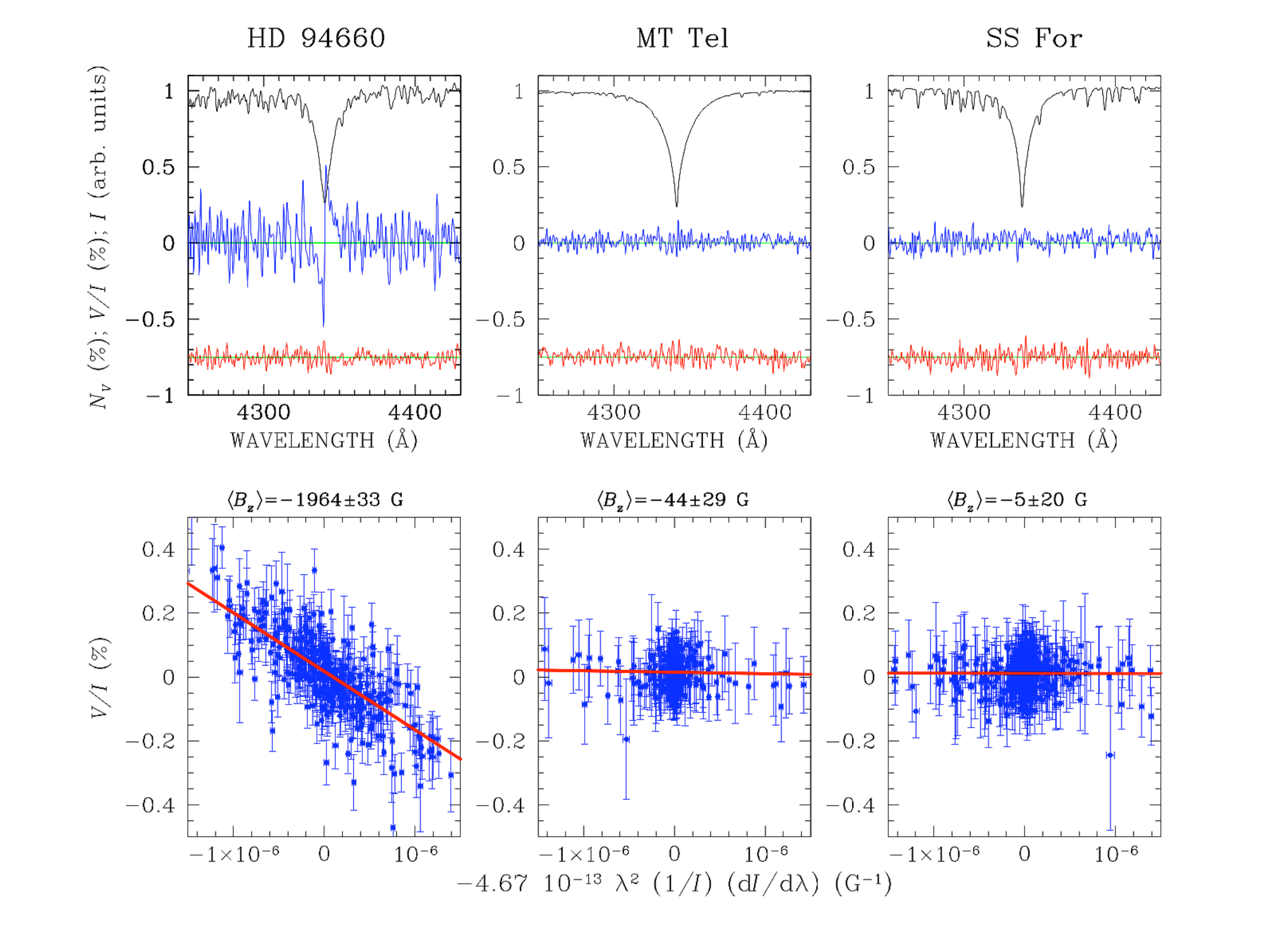}
\caption{
Spectropolarimetric observations of the magnetic star HD~94669 (left
panels) and two program RR~Lyrae stars (MT Tel, middle panels, and
SS~For, right panel). The top panels show the spectral region around
H$\gamma$. The top, solid lines represent Stokes~$I$, expressed in arbitrary
units (and not normalized to the continuum); middle lines show the
\pv\ profiles, and the bottom lines show the null profile \nv, which
represents an overal estimate of the noise of our circular polarization measurements.
The bottom panels show the corresponding best fits to the full spectrum
obtaining by minimizing the $\chi^2$ of Eq.~(\ref{Eq_ChiSquare}).
}
\label{Fig_Spectra}
\end{center}
\end{figure*}

\section{Determinations of the mean longitudinal magnetic field}
Our determinations of the mean longitudinal field \Bz\, i.e.,
the component of the magnetic field along the line of sight, averaged
over the stellar disk, are based on the use of the formula
\begin{equation}
\frac{V}{I} = - g_\mathrm{eff} \ \cz \ \lambda^{2} \
                \frac{1}{I} \
                \frac{\mathrm{d}I}{\mathrm{d}\lambda} \
                \Bz\;,
\label{Eq_Bz}
\end{equation}
where $g_\mathrm{eff}$ is the effective Land\'{e} factor and
\begin{equation}
\cz = \frac{e}{4 \pi m_\mathrm{e} c^2}
\ \ \ \ \ (\simeq 4.67 \times 10^{-13}\,\mathrm{\AA}^{-1}\ {\rm G}^{-1})\; ,
\end{equation}
where $e$ is the electron charge, $m_\mathrm{e}$ the electron mass, and
$c$ the speed of light.  

A least-squares technique was used to derive the longitudinal field
via Eq.~(\ref{Eq_Bz}), by minimizing the expression
\begin{equation}
\chi^2 = \sum_i \frac{(y_i - \Bz\,x_i - b)^2}{\sigma^2_i}
\label{Eq_ChiSquare}
\end{equation}
where, for each spectral point $i$, $y_i = (V/I)_i$, $x_i =
-g_\mathrm{eff} \cz \lambda^2_i (1/I\ \times
\mathrm{d}I/\mathrm{d}\lambda)_i$, and $b$ is a constant.

Note that Eq.~(\ref{Eq_Bz}) is strictly correct under the so-called
weak field approximation, which, e.g., in the atmosphere of main
sequence A-type stars, is valid for field strengths $\la 1$\,kG for
metal lines, and up to $\sim 10$\,kG for H lines. 

Using low resolution spectra such as those obtained in this work, it
would make sense to hypothesize that only broad H Balmer lines could
be used to determine the magnetic field, since most metal lines are
not resolved in the spectra. However, \citet{Bagetal02} and
\citet{Bagetal06} have shown that Eq.~(\ref{Eq_Bz}) may also be applied 
to spectral regions with non resolved metal lines, leading to
results consistent to the \Bz\ determinations obtained on H Balmer
lines in the weak field regime. We checked that with our spectra the
\Bz\ determination obtained from metal lines only are consistent,
within the error bars, with the \Bz\ values obtained from Balmer lines
only.  For Balmer lines we adopted an effective Land\'{e} factor of 1
\citep{CasLan94}; for metal lines, following \citet{Chaetal04}, we
used an effective Land\'{e} factor of 1.38. Finally, we applied the
least-square technique to the full spectrum, i.e., using both metal
and Balmer lines. This method allowed us to bring the \Bz\ error bars
down to a typical level of 20--30 G.

Similarly to the application by \citet{Bagetal06}, the \Bz\ values were
determined using the \pv\ profile obtained from the combinations of
all exposures. For the purpose of checking our results, we also calculated 
the average of the \Bz\ values obtained from individual pairs of
exposures for each star, always finding results consistent with the \Bz\ value
obtained from the combined \pv\ and $I$ profiles.

Since FORS1 is not equipped with circular polarization filters that
allow one to routinely check the correct alignment of the
polarimetric optics, and no standard stars for circular polarization
are available, we decided to perform a health check by
observing the well known magnetic Ap star HD~94660 (=KQ~Vel=HIC~53379) during
our observing run.  Two consecutive series of eight observations were
obtained, one with a 0.5\arcsec\ slit width, and one with a
1.0\arcsec\ slit width, using the same strategy adopted for the
program stars. HD~94660 presents a roughly constant longitudinal
magnetic field, and the measured value of $\sim -1960$\,G (from Balmer
lines only) is in reasonably good agreement with previous observations
\citep[see, e.g., Fig.~3 of][the new observation presented in this paper
would be located at phase 0.96.]{Bagetal06}.

Figure~\ref{Fig_Spectra} shows our observations of stellar spectra in
circular polarization, and their interpretation in terms of magnetic
field for HD~94660, a known magnetic star, and two representative RR
Lyrae stars: MT Tel, an RRc star, and SS For, an RRab Blazhko star.
Note that the hotter RRc star clearly has less metal lines in its
spectrum than the RRab star.  The bottom panels show the best fits to
the full spectrum obtained through minimizing the $\chi^2$ of
Eq.~(\ref{Eq_ChiSquare}).  Whereas a clear slope is observed in the
case of HD~94660, corresponding to the mean longitudinal magnetic
field \Bz\,, for the RR Lyrae stars the best fit results are consistent
with a curve with no significant slope.

\begin{figure}[ht]
\begin{center}
\resizebox{\hsize}{!}{\includegraphics{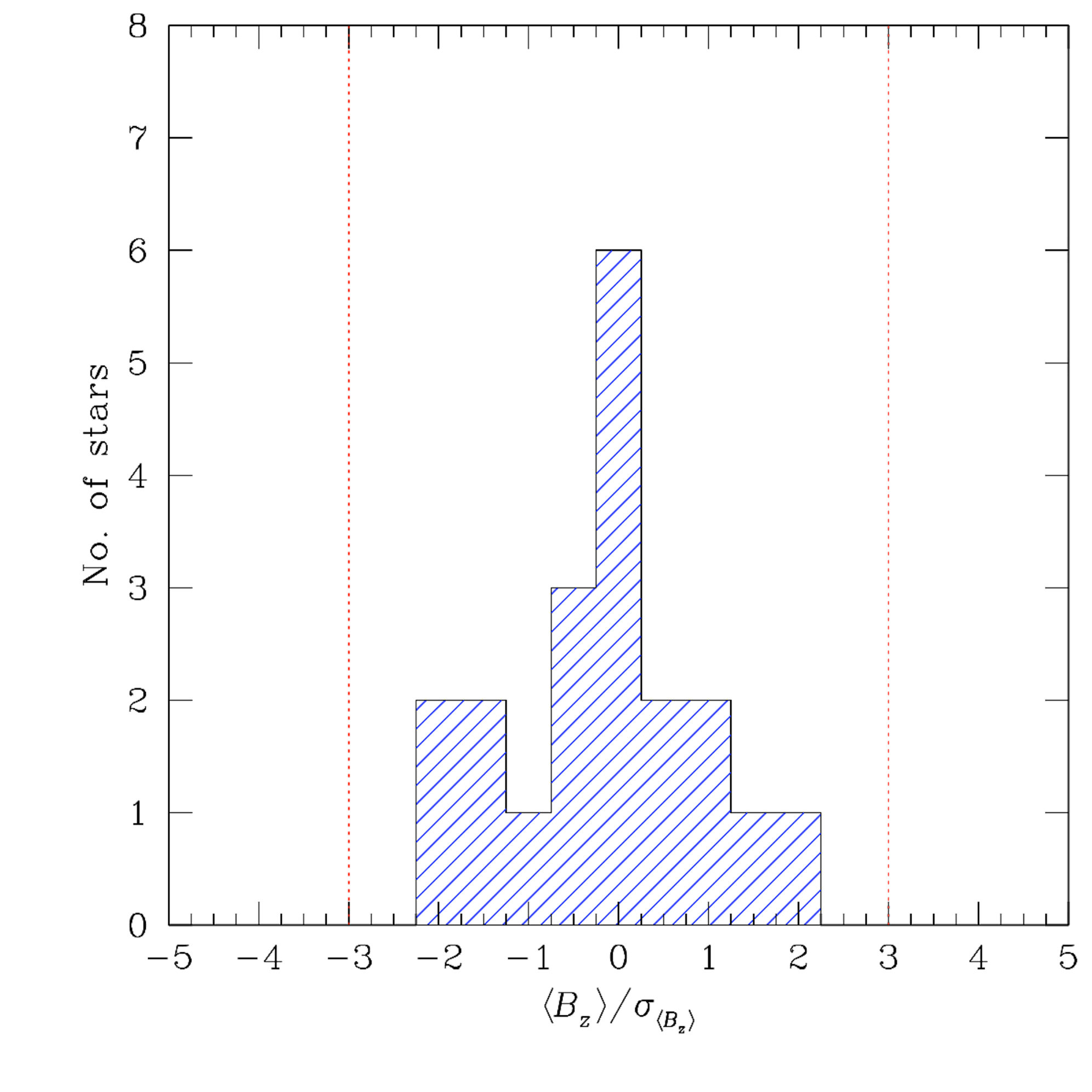}}
\caption{\label{Fig_Histogram}
Distribution of the \Bz\ values normalized to their error bars for
our sample of 20 field measurements in 17 RR~Lyrae stars. The dotted
vertical lines correspond to the 3-$\sigma$ detection level.
}
\end{center}
\end{figure}

The results of our survey are reported in
Table~\ref{Tab_Observations}.  Inspection of this table show that in
all cases our \Bz\ measurements are,  within 3$\sigma$, fully consistent with a null
detection of the mean longitudinal magnetic field.  In other words, for the 17 stars
in our sample and the 20 measurements taken, we obtained no field that
differed from zero exceeding the 3\,$\sigma$ significance level.  In
fact, 14 of our 20 measurements were null detections within a
1$\sigma$ significance level. A histogram reflecting our measurements
is given in Fig.~\ref{Fig_Histogram}.

\begin{table*}
\caption{\label{Tab_Observations} 
Determination of the mean longitudinal magnetic field of 17 RR~Lyrae
stars.  
}
\begin{center}
\begin{tabular}{llccccrcr@{\,$\pm$\,}lc}
\hline \hline
\multicolumn{1}{l}{Object}       &
\multicolumn{1}{l}{Type}         &
\multicolumn{1}{c}{Date}         &
\multicolumn{1}{c}{Time (UT)\,{\bf $^{(1)}$}}    &
\multicolumn{1}{c}{Exp\,{\bf $^{(2)}$}}          &
\multicolumn{1}{c}{Slit}         &
\multicolumn{1}{c}{$\phi$\,{\bf $^{(3)}$}}       &
\multicolumn{1}{c}{$\psi$\,{\bf $^{(4)}$}}       &
\multicolumn{2}{c}{\Bz}          &
\multicolumn{1}{c}{(SNR)\,{\bf $^{(5)}$}}          \\
\multicolumn{1}{c}{}             &
\multicolumn{1}{c}{}             &
\multicolumn{1}{c}{(yyyy mm dd)} &
\multicolumn{1}{c}{(hh:mm)}      &
\multicolumn{1}{c}{(sec)}        &
\multicolumn{1}{c}{width}        &
\multicolumn{1}{c}{($\pm$0.02)}             &
\multicolumn{1}{c}{}             &
\multicolumn{2}{c}{(G)}          &
\multicolumn{1}{c}{} \\
\hline
       &        &            &       &      &           &     & & 
\multicolumn{2}{c}{} & \\
WZ Hya & RRab   & 2008 11 12 & 08:32 & 1920 & 0.5\arcsec&0.99 &-& 32 & 21& 2620 \\ 
V Ind  & RRab   & 2008 11 13 & 03:03 & 2400 & 1.0\arcsec&0.89 &-&  $-$57& 28& 2820 \\ 
U Lep  & RRab   & 2008 11 11 & 05:06 & 2340 & 0.5\arcsec&0.23 &-&  $-$15& 19& 2930 \\ 
\vspace{2mm}
RU Scl & RRab   & 2008 11 11 & 02:26 & 3900 & 0.5\arcsec&0.63 &-&      6& 14& 2865 \\ 
CS Eri & RRc    & 2008 11 12 & 02:36 & 960  & 0.5\arcsec&0.41 &-&      3& 19& 3025 \\ 
       &        & 2008 11 13 & 06:10 & 360  & 0.5\arcsec&0.10 &-&      5& 22& 2900 \\
\vspace{2mm}
MT Tel & RRc    & 2008 11 11 & 00:10 & 1800 & 0.5\arcsec&0.07 &-&  $-$44& 29& 3370 \\ 
RV Cap & RRab-BL& 2008 11 13 & 01:56 & 4000 & 1.0\arcsec&0.17& 0.4$\pm$0.1 &   $-$4& 31& 2640 \\ 
RX Cet & RRab-BL& 2008 11 12 & 03:41 & 4560 & 0.5\arcsec&0.92 & * &     45& 21& 1970 \\ 
RV Cet & RRab-BL& 2008 11 11 & 06:03 & 2880 & 0.5\arcsec&0.31 & 0.0$\pm$0.2 &   $-$6& 15& 2535 \\ 
RX Col & RRab-BL& 2008 11 11 & 07:24 & 4800 & 1.0\arcsec&0.19 & 0.5$\pm$0.2 & $-$99& 45& 1480 \\ 
RY Col & RRab-BL& 2008 11 11 & 03:53 & 4700 & 1.0\arcsec&0.28 & 0.2$\pm$0.1 &     18& 23& 3020 \\ 
VW Dor & RRab-BL& 2008 11 12 & 07:35 & 3360 & 1.0\arcsec&0.27 &  * &  $-$38& 28& 2270 \\ 
SS For & RRab-BL& 2008 11 11 & 01:12 & 3440 & 0.5\arcsec&0.61 & 0.43$\pm$0.10 &   2& 19& 2930 \\ 
       &        & 2008 11 12 & 06:41 & 1440 & 0.5\arcsec&0.09 & 0.47$\pm$0.10 &   $-$5& 20& 2710 \\
       &        & 2008 11 13 & 04:14 & 1440 & 0.5\arcsec&0.90 & 0.49$\pm$0.10 &  $-$16& 30& 1875 \\
RX For & RRab-BL& 2008 11 13 & 05:17 & 4413 & 1.0\arcsec&0.64 & 0.8$\pm$0.4 &     15& 34& 1855 \\ 
SZ Hya & RRab-BL& 2008 11 11 & 08:40 & 1920 & 0.5\arcsec&0.16 &  * &  21& 25& 2160 \\ 
\vspace{2mm}
X Ret  &RRab-BL & 2008 11 12 & 05:16 & 4800 & 1.0\arcsec&0.76 & 0.9$\pm$0.2 &   $-$4& 29& 2250 \\ 
\vspace{2mm}
BV Aqr & RRc-BL & 2008 11 12 & 01:46 & 3360 & 1.0\arcsec&0.44 & * &      6& 27& 3620 \\ 
HD 94660& Magnetic Ap & 2008 11 12 & 09:01 &   56 & 0.5\arcsec&     & &$-$1831& 15& 3280 \\
\vspace{1mm}
       &        & 2008 11 12 & 09:08 &   28 & 1.0\arcsec&     & &$-$1872& 25 &3071 \\

\hline
\multicolumn{11}{l}{{\footnotesize $^{(1)}$ The epoch of the observations refers to the middle of the eight
exposures, and  $^{(2)}$ the exposure time refers to the sum of eight exposures.}}  \\
\multicolumn{11}{l}{{\footnotesize $^{(3)}$ $\phi$: pulsation phase;  $^{(4)}$ $\psi$: Blazhko phase. The symbol
`--' is used for non-Blazhko stars.  The symbol `*'
is used for Blazhko stars for which }} \\  
\multicolumn{11}{l}{{\footnotesize the Blazhko phase could not be obtained with an
error smaller than 0.5.}}  \\
\multicolumn{11}{l}{{\footnotesize $^{(5)}$ The
signal-to-noise ratio (SNR) is calculated per \AA, in the wavelength
interval 4475--4525\,\AA.}}  \\
\end{tabular}
\end{center}
\end{table*}

\section{Discussion}

\subsection{The magnetic model challenged}
The presence of a strong dipole magnetic field is a key ingredient for
the model proposed by \citet{ShiTak95} and \citet{Shi2000}.  Assuming
that RR Lyrae stars have fairly strong dipole fields, \citet{Shi2000}
showed that the radial mode in RR Lyrae stars would be deformed by the
Lorenz force to have additional quadrupole ($\ell = 2$) pulsation
components.  The changing aspect angle as a consequence of the
rotation of a star would then give rise to an apparent amplitude
variation.  The quadrupole components induced by the magnetic field
would give rise to a quintuplet structure in the frequency spectrum.
For more than a decade after the magnetic model was first presented,
no evidence for quintuplet components was found from any data set.
\citet{ShiTak95} showed that a quintuplet structure may manifest
itself as only a triplet depending on the geometrical configuration
(angles of pulsation axis, magnetic axis and aspect angle).  Also, the
quintuplet components might have such a low amplitude that they are
hidden in the noise of the frequency spectrum of time series data. The 
latter explanation
was supported by recent findings from high-quality data sets by Hurta
et al. (2008) in RV UMa, Jurcsik et al. (2008) in MW Lyr, and
\citet{Koletal08} in SS For, one of the stars in our sample.  However,
the detection of quintuplet structures does not imply that the magnetic
model is the one to be preferred over the resonance
model. \citet{Juretal08} and \citet{Koletal08} also found evidence of
even higher order multiplet components (septuplets, nonuplets) in the
frequency spectra of Blazhko stars.  These frequency structures have
never explicitly been considered in any of the models proposed to
explain the Blazhko effect.

Our sample of stars consisted of RR Lyrae stars with different pulsational properties: RRab, RRc stars, and modulated star of both types.  In none of these stars a field exceeding a significance level of $3\sigma$ was detected.

For the magnetic field to have the effects described by \citet{ShiTak95} and \citet{Shi2000}, 
the presence of a {\it dipole} field with a strength of the order of 1\,kG was assumed.  
In the following section we discuss what constraints our measurements set to the
magnetic field strength and morphology of our sample of stars.

\subsection{Constraints on the magnetic field}


We consider a magnetic configuration produced either by a dipole, or
by a (non-linear) qua\-dru\-po\-le, or by the superposition of both.
The model is assumed to be centered, i.e. the elementary multipoles
that produce the magnetic configuration at the stellar surface are
located at the star's center. The dipole field
$\vec{\mathcal{B}}_{\mathrm d}(\vec{r})$ can be characterized by the
dipole field strength \Bd\ at the pole, plus the direction of a unit
vector $\vec{u}$; the qua\-dru\-po\-le field $
\vec{\mathcal{B}}_{\mathrm q}(\vec{r})$ can be characterized by the
quadupole field strength \Bq, plus the directions of two unit vectors
$\vec{u}_1$ and $\vec{u}_2$.  The geometric configuration of a dipole
plus quadrupole magnetic field is illustrated in Fig.~1 of
\citet{Lanetal98}.  Denoting by $R_*$ the star's radius, the magnetic
field vector at a given point $\vec r$ of the stellar surface is
given, respectively, by
\begin{equation}
\begin{array}{rcl}
 \vec{\mathcal{B}}_{\mathrm d}(\vec{r})
 &=&-\frac{\Bd}{2}
    \left[\,\vec{u}-3\,\frac{\vec{u}\cdot\vec{r}}{R_*^2}\;\vec{r}\,\right]
\label{Eq_Dip}\\[0.15cm]
 \vec{\mathcal{B}}_{\mathrm q}(\vec{r})
 &=&-\frac{B_{\mathrm{q}}}{2}
    \left[\frac{\vec{u_2}\cdot\vec{r}}{R_*}\;\vec{u_1}\,+\,
	  \frac{\vec{u_1}\cdot\vec{r}}{R_*}\;\vec{u_2}\right.
							  \nonumber\\[0.15cm]
 & &\hphantom{-\frac{\Bq}{2}\biggl[}
   \left.+\left(\frac{\vec{u_1}\cdot\vec{u_2}}{R_*}-
		5\,\frac{(\vec{u_1}\cdot\vec{r})(\vec{u_2}\cdot\vec{r})}
		      {R_*^3}
	  \right)\vec{r}\,\right]\ .                        
\label{Eq_Bqua}
\end{array}
\end{equation}
Note that while the quantity \Bd, the dipole field modulus at the
magnetic pole, has an obvious physical meaning, the quantity \Bq\
represents the qua\-dru\-po\-le field modulus at the magnetic pole
\textit{only} when the unit vectors $\vec{u}_1$ and $\vec{u}_2$
coincide (linear qua\-dru\-po\-le), otherwise the concept itself of
magnetic pole becomes meaningless, and the estimate of the actual
field strength requires an evaluation of Eq.~(\ref{Eq_Bqua}) for the
specific quadrupole configuration. For a more detailed description of
the multipolar expansion of the magnetic field see \citet{Bagetal96}.

Assuming a limb-darkening law of the form
\begin{equation}
1 - u + u\,\mu\ \ (0\le u\le 1) \;,
\end{equation}
where $\mu$ is the cosine of the angle between the line of sight through the
star disk center and a given point of the stellar disk, and using Eqs.~(24), 
(28) and (68) of \citet{Bagetal96}, we find that the mean longitudinal field is
given by 
\begin{equation}
\begin{array}{rcl}
\Bz &=& \frac{3}{3-u}\,\Big[\frac{15\,+\,u}{60} \Bd \cos \ell \\
&+& 
                           \frac{u}{12}         
\Bq \big(\cos \ell_1 \cos \ell_2  + \frac{1}{2} 
         \sin \ell_1 \sin \ell_2 \cos(\Lambda_1 - \Lambda_2)\big) \Big]\\
\end{array}
\label{Eq_Bzdip+Bzqua}
\end{equation}
where $\ell$ is the angle between the line of sight and the dipole
axis, $\ell_1$ and $\ell_2$ are the angles between line of sight
and the unit vectors $u_1$ and $u_2$ that define the quadrupolar
component of the magnetic field, and $\Lambda_1$ and $\Lambda_2$ are
the azimuth angles of the same unit vectors, reckoned
counterclockwise in the plane of the sky from the north celestial
meridian, as detailed in \citet{Bagetal96}.

Even a quick glance at the results in Table~\ref{Tab_Observations}
tells us that, if a dipole field is present, its strength must be
small, much smaller than the 1\,kG field that the magnetic model
\citep{Shi2000} requires to be responsible for the Blazhko effect.
Based on our measurements, it is possible to obtain more precise
constraints on the magnetic field of our program stars.

Let us consider the case of a pure dipolar field. We assume that all
stars of our sample have the same dipolar strength, and that the
magnetic axis is randomly oriented with respect to the line of sight.
The probability $P(\ell)$ for the tilt angle of the
field $\ell$ to be in the range $[\ell,\ell+\mathrm{d}\ell]$ is proportional to
$\sin\ell \mathrm{d}\ell$.  A simple application of Bayesian
statistics allows us to evaluate the probability $\mathcal{P}(\Bd \ge \Bdo)$
that all the observed
stars of our sample are characterized by a dipolar field strength
higher than a certain threshold value \Bdo:
\begin{equation}
\begin{array}{l}
\mathcal{P}(\Bd \ge \Bdo) \propto \\
\ \ \ \
\int\limits_{\Bdo}^{+\infty}\,\mathrm{d} \Bd \ \
\prod\limits_{j=1}\limits^{N}\
\frac{1}{\sqrt{2 \pi} \sigma_j} \,
\int\limits_{0}^{\pi}\,     \mathrm{d}\ell
\sin \ell \,
\exp{\left(-\frac{\big(\Bz_j - k(u) \Bd \cos \ell \big)^2}{2\,\sigma^2_j}\right)} \\
\end{array}
\end{equation}
where the product runs on the $N=17$ observed stars (for the \Bz\
value of the two stars that were observed more than once, we consider the value 
measured first), 
$\sigma_j$ is the error bar associated to the measurement of the $j^{\rm th}$ star, 
and we have set
\begin{equation}
k(u) = \frac{15 + u}{20\, (3-u)} \ .
\end{equation}
Using a limb-darkening coefficient $u=0.62$, derived from
\citet{DiaCor95} \citep[see also][]{Kolenberg02}, we obtain that if
the RR~Lyrae stars of our sample are all characterized by the same dipolar
strength, with magnetic axis randomly oriented about the line
of sight, then there is a 95\,\% probability that their dipolar
strength is $\la 130$\,G.

Equation~(\ref{Eq_Bzdip+Bzqua}) shows that, for the limb-darkening
coefficient of $u=0.62$, the contribution of the quadrupolar component
to the mean longitudinal magnetic field is always $\la 20$\,\% the
contribution due to the dipolar component (assuming \Bd\ = \Bq). For a
linear quadrupole, i.e., with $\vec{u_1} \parallel\vec{u_2}$,
we always have $\ell_1 = \ell_2$ and $\Lambda_1 = \Lambda_2$, and our observations
rule out a quadrupolar field strength
$\ga 250$\,G at the 95\,\% confidence level. If we consider the special case where $\ell_2 = \ell_1 + 90\degr$ and $\Lambda_1 = \Lambda_2$ (for which $\vec{u_1} \perp \vec{u_2}$), we obtain the value \Bq = 1.9 kG  as an upper limit to the quadrupole strength.

Therefore, determinations of the mean longitudinal field as obtained
in this work set fairly loose constraints on the magnetic field
strength for morphologies of higher order than the dipole field.

\subsection{Variability over pulsation cycle and Blazhko cycle}
\begin{figure}[ht]
\begin{center}
\resizebox{\hsize}{!}{\includegraphics{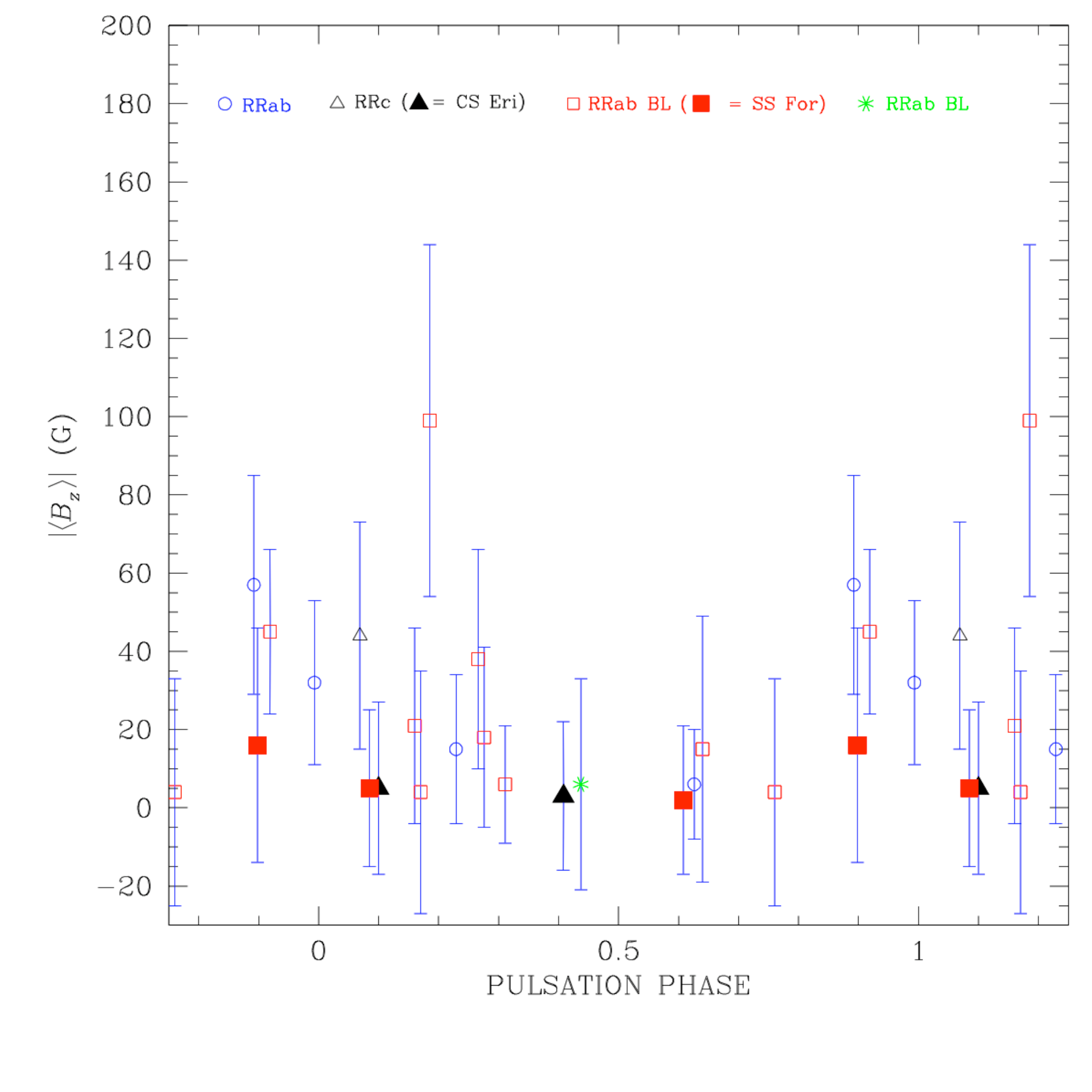}}
\caption{\label{Fig_Phase}
Absolute value of the mean longitudinal field versus pulsation phase.
Different symbols were used for different types of RR Lyrae stars.
Data points for stars observed more than once are denoted with
filled symbols.
}
\end{center}
\end{figure}
Figure~\ref{Fig_Phase} shows the plot of the absolute value of the
mean longitudinal field as a function of the pulsation phase. Although
this plot refers to different stars with different geometric
configurations, it could potentially reveal the presence of a cyclic
variability of the magnetic field {\it over the
pulsation cycle} as was also observed on several nights by \citet{Rometal87} 
(their Figs.\,1 and 2). Indeed, Figure~\ref{Fig_Phase} shows a hint for higher \Bz\ values (though
all consistent with zero) close to phase zero.  This could be
explained by the presence of a small magnetic field and assuming
magnetic flux conservation. (The stellar radius and surface is minimum shortly before phase
zero, and attains its highest value around phase $\phi=0.5$.) Further conclusions may be drawn
only after a detailed monitoring of individual stars.

For two of the stars in our sample, CS Eri (RRc) and SS For (RRab-BL), we obtained more than one
measurement at different phases in their pulsation cycle.  No clear variation with the pulsation phase 
exceeding the error bars could be derived for these stars.
We observed SS For at 3 different phases: one at maximum
light, a second one approaching minimum light, and a third one on the
rising branch of the light variation.  From the last spectrum a higher absolute \Bz\ was derived, but also 
a higher error bar, which can be attributed to the lower signal-to-noise ratio of the spectrum.
SS For shows quintuplets in its
pulsation spectrum \citep{Koletal08} which, according to the magnetic model \citep{Shi2000} are associated with the
effects of a magnetic field (see also Section 4.1), and hence was one
of our prime targets for investigating the presence of a magnetic
field.

\citet{Rometal87} also reported a variation of their derived mean longitudinal magnetic field values {\it over the Blazhko cycle} of RR Lyr. 
\citet{Chaetal04}, in contrast, with 27 measurements of the same star, detected no significant changes at different Blazhko phases.
Within the allocated telescope time we were not able to follow one Blazhko star at different Blazhko phases.  Only for some stars in our sample the Blazhko phase could be determined with a reasonable accuracy. However, we can safely say that the 11 Blazhko stars 
in our sample were recorded at random Blazhko phases, and for all of them we obtained a mean longitudinal magnetic field consistent with a null detection within $3\sigma$.  As discussed in Section~4.2, this implies low values for the magnetic dipole field.

Our observations significantly enlarge the sample of spectropolarimetric measurements of RR Lyrae stars and follow the line of results obtained by \citet{Preston67} and \citet{Chaetal04}. The latter authors declared RR Lyr to be a bona fide non-magnetic star on the basis of high-resolution spectropolarimetric data obtained over a 4-year time span.

Now why did earlier measurements of the magnetic field in specifically RR Lyr, the brightest star of the class, lead to reported values of the longitudinal magnetic field strength as high as about 1.5\,kG \citep{Babcock58,Rometal87}, while more recent observations contradict this?
\citet{Chaetal04} addressed this question in the necessary detail (in their Section\,3.1), pointing at the limitations of the use of photographic plates resulting in underestimated error bars, and the distortions of the spectral lines caused by shock waves in certain pulsation phases \citep[see also][]{Wadetal02}.  The measurement with by far the highest \Bz\ detection by \citet{Rometal87} was recorded at the phase of the main shock in the star ($\phi$=0.95).  This may have led to spurious polarization signals. We also note two limiting factors of those earlier measurements: i) a quite low reciprocal dispersion (9\AA/mm) of their plate material, which is possibly not sufficient to perform high accuracy splitting measurements of the spectral lines observed in opposite polarization; ii) the employment of the Nasmyth focus station for the polarimeter: the oblique reflection from the tertiary mirror is prone to introduce a spurious polarization signal. This problem is much less critical at the Cassegrain focus, which was used instead for both the MuSiCoS observations obtained by \citet{Chaetal04}, and the FORS1 measurements presented in this work. We therefore suggest that the magnetic field in RR Lyr was significantly overestimated in the earlier studies.

\section{Conclusion}

The results of our survey of magnetic fields in RR Lyrae stars reveal a serious challenge to the magnetic models for explaining the Blazhko effect.  
For a sample consisting of 17 RR Lyrae stars with different pulsational properties, no substantial dipole magnetic field, as required by the magnetic model \citep{Shi2000}, was detected.  
We determined an upper limit for the strength of the dipole field in the stars of our sample at 130\,G with a 95\% confidence level.
Our result implies that the Blazhko modulation in
the pulsation of RR Lyrae stars is not correlated with the presence of
a strong, quasi-dipolar magnetic field.
For morphologies of higher order than the dipole field, however, our determinations of the mean longitudinal field as obtained
in this work only set fairly loose constraints on the magnetic field strength. More complex magnetic field morphologies may be detected with high-resolution spectropolarimetric data.


\begin{acknowledgements}
This paper is based on observations made with ESO Telescopes at the
La Silla-Paranal Observatory under program ID 082.D-0342.
KK is supported by the Austrian Fonds zur 
F\"orderung der wissenschaftlichen Forschung, project number P19962-N16 and T359-N16.
SB thanks D.J.~Asher for very useful discussions.  We thank the referee for constructive comments.
The research has made use of the SIMBAD astronomical database 
(http://simbad.u-strasbg.fr/) and the GEOS RR Lyrae database 
(http://dbrr.ast.obs-mip.fr/). 
\end{acknowledgements}

\end{document}